\documentclass[12pt]{article} 
\usepackage{graphicx} 

\def \bea{\begin{eqnarray}} 
\def \beq{\begin{equation}}

\def \eea{\end{eqnarray}} 
\def \eeq{\end{equation}}

\def \3half{\frac{3}{2}}

\begin{document} 
\begin{flushright} 
TECHNION-PH-2015-11\\
EFI 15-29\\ 
August 2015 \\ 
\end{flushright} 
\centerline{\bf Comment on ``A possible explanation of the}
\vskip 1mm
\centerline{\bf D0 like-sign dimuon charge asymmetry"} 
\bigskip 
\centerline{Michael Gronau} 
\medskip 
\centerline{\it Physics Department, Technion - Israel Institute of Technology} 
\centerline{\it Haifa 32000, Israel} 
\medskip 
\centerline{Jonathan L. Rosner} 
\medskip 
\centerline{\it Enrico Fermi Institute and Department of Physics, 
 University of Chicago} 
\centerline{\it Chicago, IL 60637, U.S.A.} 
\bigskip 
\begin{quote} 
We show that a contribution due to a second order amplitude with intermediate $\bar u d$ 
in a loop, which was claimed by Descotes-Genon and Kamenik to dominate the CP 
asymmetry in $b \to c \ell \nu$, vanishes.
\end{quote} 
 \bigskip
 
In a 2013 paper by S. Descotes-Genon and J. F. Kamenik~\cite{DescotesGenon:2012kr} 
(discussing the D0 like-sign dimuon asymmetry \cite{Abazov:2010hv})  the authors 
presented a Standard Model calculation of a contribution claimed to dominate the 
direct CP asymmetry $A^{b~{\rm SM}}_{\rm dir}$ in inclusive semileptonic decays 
$b \to c \ell \nu~(\ell=\mu)$. Their result was stated to be an order of magnitude 
larger than a value, $A^{b~{\rm SM}}_{\rm dir}\equiv A^b_{sl} =-3.2 \pm 0.9)\times 10^{-9}$
calculated by us in collaboration with S. Bar-Shalom and G. Eilam  
\cite{BarShalom:2010qr}. In this brief comment we wish to clarify this point of discrepancy.

\begin{figure}[h] 
\begin{center} 
\includegraphics[height=2.8cm]{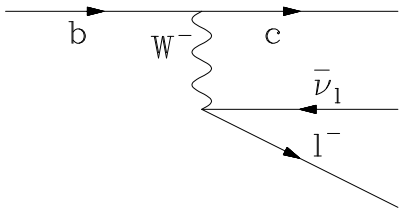}  
\end{center} 
\caption{Tree diagram for $b \to c \ell^- \bar\nu_{\ell}$ 
\label{fig:tree}} 
\end{figure}

As argued in Ref.~\cite{BarShalom:2010qr} using CPT, a nonzero asymmetry in 
$b \to c \ell\nu$ requires interference of a tree level amplitude described in 
Fig.\ \ref{fig:tree} with an amplitude which is second order in weak interactions. 
%
%
\begin{figure}[h] 
\begin{center} 
\includegraphics[height=2.8cm]{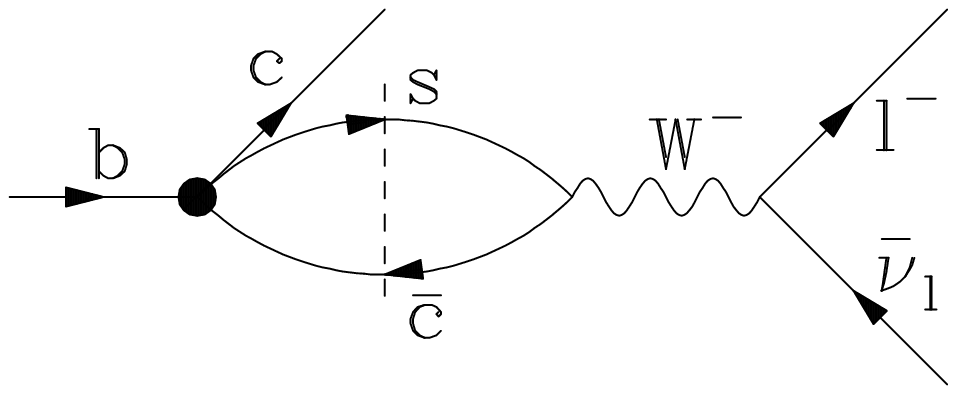} 
\end{center}
\caption{Second-order diagram for $b \to c \ell^- \bar\nu_{\ell}$.  
\label{fig:loop}} 
\end{figure} 
In order to produce an asymmetry, the second amplitude must 
involve a CKM factor with a {\em different weak phase} and a nonzero CP-conserving
phase. A second-order amplitude fulfilling these two requirements is drawn 
in Fig.\,\ref{fig:loop}, consisting of a product 
of a penguin amplitude for $\bar b \to \bar c c\bar s$ involving $V^*_{tb}V_{ts}$ 
and a tree amplitude for $c \bar s \to \ell^+ \nu_{\ell}$ 
involving $V^*_{cs}$. A relative CP-conserving phase of $90^\circ$ 
between the two amplitudes follows by taking the absorptive 
part of the second-order amplitude. The absorptive part is described by a 
discontinuity cut 
crossing the $\bar c  s$ lines in the second-order diagram, which amounts 
to summing over corresponding on-shell intermediate states. A detailed calculation,
using a value for the weak phase difference between the two amplitudes~\cite{CKMf},
${\rm Arg}\left(V_{tb} V_{ts}^\star V_{cs} V_{cb}^\star \right) \equiv \beta_s = 0.018$, 
and including uncertainties in $b$ and $c$ quark masses, led to the above-mentioned asymmetry result.  

Ref.~\cite{DescotesGenon:2012kr} proposed an alternative mechanism 
claimed to dominate the asymmetry, replacing the intermediate $\bar c s$ in 
Fig.\,\ref{fig:loop} by intermediate $\bar u d$ coupled by a tree amplitude for 
$b \to c \bar u d$. Interference of this second order amplitude with the tree 
amplitude in Fig.\ \ref{fig:tree} was stated to involve a weak phase factor 
${\rm Im}\left(V_{ub} V^\star_{ud} V_{cd} V^\star_{cb}\right)$ 
[see Eqs.\,(21a) and (22) in \cite{DescotesGenon:2012kr}]. This factor 
would seem to describe a second order amplitude involving intermediate 
$\bar d u$ which violates charge conservation.  The
actual imaginary part of the CKM factor for intermediate $\bar u d$ vanishes:
\beq
{\rm Im}\left(V_{cb} V_{ud}^\star V_{ud} V_{cb}^\star \right) = 0~.
\eeq
Thus this interference term vanishes and does not contribute at all to the 
asymmetry. 
\bigskip

M.G. is grateful to the CERN department of theoretical physics for its 
hospitality and wishes to thank Jernej Kamenik for a discussion.
The work of J.L.R. is supported in part by the United States Department of
Energy, Division of High Energy Physics, Grant No.\ DE-FG02-13ER41958.
  
\end{document}